\documentclass[aps,preprint,nofootinbib]{revtex4}
\begin{document}
\title{\Large \bf  Rotating Black Holes in Higher Dimensional Einstein-Maxwell Gravity}
\date{\today}
\author{\large A. N. Aliev}
\affiliation{TUBITAK-BU,  Feza G\"ursey Institute, P.K. 6  \c
Cengelk\" oy, 81220 Istanbul, Turkey}

\begin{abstract}

The strategy of obtaining the familiar Kerr-Newman solution in
general relativity  is based on either using the metric ansatz in
the Kerr-Schild form, or applying the method of complex coordinate
transformation to a non-rotating charged black hole. In practice,
this amounts to an appropriate re-scaling of the mass parameter in
the metric of uncharged black holes. Using a similar approach, we
assume a special metric ansatz in $\,N+1\, $ dimensions and
present a new analytic solution to the Einstein-Maxwell system of
equations. It describes rotating charged black holes with a single
angular momentum in the limit of slow rotation. We also give the
metric for a slowly rotating charged black hole with two
independent angular momenta in five dimensions. Finally, we
compute the gyromagnetic ratio of these black holes which
corresponds to the value $\,g=N-1\,$.

\vspace{5mm} PACS number(s):  04.70.Bw, 05.50.+h

\end{abstract}
\maketitle
\section{Introduction}

Black holes were originally predicted in the framework of four
dimensional general relativity as the endpoint of gravitational
collapse of sufficiently massive stars. Subsequently apart from
their astrophysical implications, they have also played a profound
role in understanding the nature of general relativity itself,
resulting in the famous singularity theorems \cite{penrose, hp}.
It has been shown that black holes possess a number of remarkable
features such as the equilibrium and uniqueness properties and
quantum properties of evaporation of microscopic black holes
\cite{Israel,hawking1,hawking2}.

From a pure theoretical point of view one can expect that the
properties of black holes might also have played an important role
in understanding the nature of gravity in higher dimensions. This
expectation has triggered the study of black hole solutions in
higher-dimensional gravity theories as well as in string/M-theory
\cite{gm, hs} (see Refs.\cite{youm} for reviews). Developments
have revealed both new possibilities to test the predictions of
string/M-theory and new unexpected features of black holes
inherent in higher dimensions. For instance, by counting the
microstates of certain supersymmetric black holes in five
dimensions it has become possible to explain the statistical
origin of the Bekenstein-Hawking entropy \cite{bmpv}. Furthermore,
it turned out that the higher dimensions allow different horizon
topologies for black holes, whereas in four dimensions the event
horizon is uniquely determined by the topology of a two-sphere
\cite{Israel, hawking1}. Accordingly, some basic properties of
black holes change when going from four dimensions to higher
dimensions, among them are {\it stability} and {\it uniqueness}
properties. The simplest class of extended black holes (black
strings with cylindrical topology of the horizon) exhibits the
linear perturbative instability below a certain critical mass
\cite{glaf}. There exists a strong evidence that the endpoint of
this instability may result in a separate black hole, thereby
providing an example of non-uniqueness in the form of a phase
transition between black holes and black strings in higher
dimensions \cite{kol}.

The first higher dimensional black hole solutions with the
spherical topology of the horizon have been found in \cite{tang}.
These solutions generalize the familiar spherically symmetric
Schwarzschild and Reissner-Nordstrom solutions of four-dimensional
general relativity. It is remarkable that for the static black
holes the uniqueness and the stability properties still survive
\cite{gibbons1} in higher dimensions, however the situation is
drastically changed for rotating black holes. The rotating black
hole solution was found by Myers and Perry \cite{mp}. This
solution is not unique, unlike its four dimensional counterpart,
the Kerr solution. There exists a rotating black ring solution in
five dimensions with the horizon topology of $\; S^2 \times S^1\,
\,$ \cite{er} which may have the same mass and angular momentum as
the Myers-Perry solution. Thus, the black ring solution can be
thought of as describing  a "donut-shaped" rotating black hole
which is absent in four dimensions. This provides another example
of a lack of uniqueness for black holes in higher dimensions. As
for the stability of these solutions, it is still not known in the
general case, though it has been argued that the Myers-Perry
solution becomes unstable at large enough rotation for a fixed
mass \cite{em}. The different physical properties of black holes
and black rings  in higher dimensions have been discussed in a
number of papers \cite{fs1}-\cite{maeda}.

After the advent of Large Extra Dimension Scenarios
\cite{ADD}-\cite{RS1} the study of black holes in higher
dimensions got a new strong impetus. These scenarios consider our
observable Universe as a slice,- a "3-brane" in higher dimensional
space and give  an elegant geometric resolution of the hierarchy
problem between the electroweak scale and the fundamental scale of
quantum gravity. The large size of the extra dimensions supports
the weakness of Newtonian gravity on the brane and makes it
possible to lower the scale of quantum gravity down to the same
order as the electroweak interaction scale (of the order of a few
TeVs). One may naturally assume that in such scenarios black holes
would be mainly localized on the brane, however some portion of
their event horizon would have a finite extension into the extra
dimension as well. On the other hand, it is clear that if the
radius of the horizon of a black hole on the brane is much smaller
than the size scale of the extra dimensions $(\,r_{+} \ll L \,)$,
the black hole would behave as a higher dimensional object. These
black holes, to a good enough approximation, can be well described
by the exact solutions of higher dimensional Einstein's equations
\cite{tang, mp}. Thus, the Large Extra Dimension Scenarios open up
new exciting possibilities to relate the properties of higher
dimensional black holes to the observable world by direct probing
of TeV-size mini black holes at future high energy colliders
\cite{gt}.

In  the light of all described above it becomes obvious that the
further study of black hole solutions in higher dimensional
gravity is of great importance. Of particular interest is the case
of charged black holes as after all, black holes produced at
colliders may, in general, have an electric charge as well as
other type of charges. Though the non-rotating black hole solution
to the higher dimensional Einstein-Maxwell equations was found a
long time ago \cite{tang}, rotating charged black holes have been
basically discussed in the framework of certain supergravity
theories and string theory \cite{town, youm, cvetic1}. The
rotating black hole solution in higher dimensional
Einstein-Maxwell gravity, that is the counterpart of the usual
Kerr-Newman solution, still remains to be found analytically.
Numerical solutions for some special cases in five dimensions have
been found in \cite{knp}.

In this paper we shall present new analytical solutions to the
higher dimensional Einstein-Maxwell equations which describe
electrically charged black holes with slow rotation. The
organization of the paper is as follows. In Sec.II we begin with a
brief description of the Myers-Perry metric with a single
non-vanishing rotation parameter in $ N+1$ dimensions. We describe
the Killing isometries of the metric, its mass parameter and
specific angular momentum. Assuming that the Myers-Perry black
hole may also have a small electric charge we construct the
potential one-form for the electromagnetic field of the charge. In
Sec.III we assume a special ansatz for the metric of charged
rotating black holes following the strategy of obtaining the
consistent solution to the Einstein-Maxwell system of equations in
four dimensional space-time. With this ansatz we show that the
simultaneous solution of the Einstein-Maxwell equations exists
only when the rotation of the black hole occurs slowly. Next, in
Sec.IV we extend this approach to examine the metric of a slowly
rotating charged black hole with two independent angular momenta
associated with two orthogonal $2$-planes of rotation in five
dimensions. Finally, in Sec.V  we compute the value of the
gyromagnetic ratio for rotating charged black holes in $\,N+1\,$
dimensions which is turned out to be equal to $\,g=N-1\,$. In
Appendix A we present the non-vanishing components of the
electromagnetic source tensor in arbitrary spacetime dimensions.
The components of the Ricci tensor for given metric ansatz are
calculated  in Appendix B.

\section{The Myers-Perry black hole with a weak electric charge}

\subsection{Properties of the metric}

The general stationary and asymptotically flat metric describing
rotating black holes with multiple angular momenta in different
orthogonal planes of $ N+1 $ dimensional space-time was found by
Myers and Perry \cite{mp}. We recall that in $ N+1 $ dimensions
the rotation group is $  SO(N) $ which possesses $\lfloor N/2
\rfloor$ independent Casimir invariants. (The notation $\lfloor
N/2 \rfloor$ denotes the integer part of $ N/2$). To be precise,
the angular momentum of the system in general is described by an
anti-symmetric $ N\times N $ tensor. In the center-of-mass frame
one can transform this tensor into its block-diagonal form by a
suitable rotation of the spatial coordinates. In this case the
angular momentum tensor is characterized by $\lfloor N/2 \rfloor$
physical parameters which in accordance with the existence of
$\lfloor N/2 \rfloor$ Casimir invariants are associated with
rotations in distinct planes. We shall focus on higher-dimensional
black holes with a single rotation parameter. The corresponding
metric in the Boyer-Lindquist type coordinates can be written in
the form
\begin{eqnarray}
ds^2 & = &- \left(1-\frac{m}{r^{N-4} \,\,\Sigma}\,\right) dt^2+
{{r^{\,N-2}\, \Sigma }\over \Delta}\,dr^2+ \Sigma \, d\theta^2  +
\left(r^2+a^2+ \frac{m  a^2 \sin^2\theta}
{r^{N-4}\,\Sigma}\right)\sin^2{\theta}\,d\phi^2 \nonumber
\\[3mm] && - {\frac{2 m  a \sin^2 \theta}{r^{\,N-4}\,\Sigma}}\,dt \,d\phi+ r^2 \cos^2{\theta}
\, d\Omega_{N-3}^2\,\,, \label{metric}
\end{eqnarray}
where
\begin{equation}
\Sigma=r^2+a^2 \,\cos^2\theta  \;,\;\;\;\;\;\; \Delta= r^{N-2}(r^2 +
a^2) -m \, r^2 \,\,,
\end{equation}
the parameter $\,m \,$  is related to the mass of the black hole,
while  $\,a \,$  is a parameter associated with its angular momentum
and
\begin{equation}
d\Omega_{N-3}^2 =d {\chi_{1}}^2+ \sin^2{\chi_{1}}\,(\,d
{\chi_{2}}^2+\sin^2{\chi_{2}}\,(...d{\chi_{N-3}}^2...)\,)
\label{sphmetric}
\end{equation}
is the metric of a unit $(N-3)$-sphere. The square root of the
determinant of the metric (\ref{metric}) is given by
\begin{equation}
\sqrt{-g}= \sqrt{\gamma}\,\,\Sigma\,r^{\,N-3} \sin{\theta}\,
\cos^{\,N-3}{\theta}\,\, \label{determinant}
\end{equation}
where $\gamma$ is the determinant of the metric (\ref{sphmetric}).
When dropping the last term the metric (\ref{metric}) bears a
close resemblance to its counterpart, the Kerr solution of
ordinary general relativity, exactly covering it for $N=3$.

The event horizon is a null surface determined by the equation
$\,\Delta=0\,$,  which implies that
\begin{eqnarray}
r^2 + a^2 - \frac{m}{r^{N-4}}\, =0 \,\,. \label{nullsurf}
\end{eqnarray}
The largest real root of this equation gives the location of the
black hole's outer event horizon. We see that the properties of the
horizon essentially depend on the dimension of the space. In
particular, from equation (\ref{nullsurf}) it follows that in $N=3$
and $N=4$ dimensions the event horizon exists  until its rotation
attains the maximum speed bounded by the mass of the black hole.
However, for $N \geq 5$ the horizon does exist independently of the
rotation, that is the black hole with a given mass  may have
arbitrarily large angular momentum \cite{mp,em}.

The time-translation invariance of the metric (\ref{metric}) along
with its rotational  symmetry in the $\,\phi\,$- direction imply
the existence of the commuting Killing vectors
\begin{equation}
{\bf \xi}_{(0)}=\xi^{\mu}_{(t)} \,\frac{\partial}{\partial
x^{\mu}}\,, ~~~~~~~{\bf \xi}_{(3)}=\xi^{\mu}_{(\phi)}
\,\frac{\partial}{\partial x^{\mu}}\,,
  \label{killing}
\end{equation}
such that their various scalar products are expressed through the
metric components as follows
\begin{eqnarray}
{\bf \xi}_{(0)} \cdot {\bf \xi}_{(0)}&=& g_{00}= -1+
\frac{m}{r^{N-4}\Sigma}\,\,,\nonumber \\[3mm]
{\bf \xi}_{(0)} \cdot {\bf \xi}_{(3)}&= &g_{03}= -\,\frac{m\,a
\sin^2 \theta}{r^{N-4} \Sigma}\,\,,
\nonumber \\[3mm]
{\bf \xi}_{(3)} \cdot {\bf \xi}_{(3)}&=& g_{33}= \left(r^2+a^2 +
\frac{m\,a^2 \sin^2\theta}{r^{N-4}
\Sigma}\right)\,\sin^2\theta\,\,. \label{kproduct}
\end{eqnarray}

The Killing vectors (\ref{killing}) can be used to give a physical
interpretation of the parameters $\,m\,$\,, $\,a\,$  in the
metric. Indeed, using the analysis given in \cite{komar}, we can
obtain the following coordinate-independent definitions for these
parameters
\begin{eqnarray}
m=\frac{1}{(N-2)\, A_{N-1}}\,\oint \,\xi_{(t)}^{\mu\,;\,\nu}\,
d^{N-1} \Sigma_{\mu\;\nu}\,\,, \label{mass}
\end{eqnarray}
and
\begin{eqnarray}
j & =&a\,m=-\,\frac{1}{2\,A_{N-1}}\,\oint
\,\xi_{(\phi)}^{\mu\,;\,\nu}\, d^{N-1} \Sigma_{\mu\;\nu}\,\,.
\label{momentum}
\end{eqnarray}
The semicolon denotes covariant differentiation and the integrals
are taken over the $(N-1)$-sphere at spatial infinity with the
surface element
\begin{equation}
d^{N-1} \Sigma_{\,\mu\,\nu}= \frac
{1}{(N-1)\,!}\,\sqrt{-g}\,\,\epsilon_{\,\mu\,\nu\,i_{1}\,i_{2}\,....\,i_{N-1}}\,
dx^{i_{1}}\wedge dx^{i_{2}}\wedge...\wedge dx^{i_{N-1}}\,\,,
\end{equation}
the quantity
\begin{equation}
A_{N-1}= \frac{2\,\pi^{\,N/2}}{\Gamma(N/2)}\,\,
\end{equation}
gives the area of a  unit $(N-1)$-sphere. We have introduced the
specific angular momentum parameter $\,j \,$ associated with
rotation in the $\,\phi\,$ direction. We note that with these
definitions the relation between the specific angular momentum and
the mass parameter looks exactly like the corresponding relation
$\,(J=a\,M) \,$ of four dimensional Kerr metric.

To justify the definitions  (\ref{mass}) and (\ref{momentum}) we
can calculate the integrands in the asymptotic region $\,r
\rightarrow \infty\,$. For the dominant terms in the asymptotic
expansion  we have
\begin{eqnarray}
\xi_{(t)}^{t\,;\,r}& = & \frac{m\, (N-2)}{2\,r^{N-1}} +
\mathcal{O}\left(\frac{1}{r^{N+1}}\right)\,\,,
\nonumber \\[4mm]
\xi_{(\phi)}^{t\,;\,r}& = & - \,\frac{j\, N
\sin^2\theta}{2\,r^{N-1}}+
\mathcal{O}\left(\frac{1}{r^{N+1}}\right)\,\,. \label{mj}
\end{eqnarray}
One can easily show that the substitution of these expressions
into the formulas (\ref{mass}) and (\ref{momentum}) verifies them.
The relation of the above parameters to the total mass $\,M\,$ and
the total angular momentum $\,J \,$ of the black hole can be
established using the standard Komar integrals in N+1 dimensions
\cite{mp}. We find that
\begin{eqnarray}
m ={\frac{16 \pi G}{N-1}}\,{\frac{M}{A_{\,N-1}}}\,\,,~~~~~~~~j =
\frac{8 \pi G J}{A_{\,N-1}}\,\,. \label{total}
\end{eqnarray}
These relations confirm the interpretation of the parameters
$\,m\,$ and  $\,a\,$  as being related to the physical mass and
angular momentum of the black hole .

\subsection{Electromagnetic potential one-form}

Let us now assume that a rotating black hole in $\,N+1\,$
dimensions possesses an electric charge $\,Q\,$. When the charge
is small enough compared with the mass of the black hole, $\,Q\ll
M\,$, the influence of the electromagnetic field of the charge on
the metric of space-time may become negligible, that is the
space-time can still be well described by the Myers-Perry metric
(\ref{metric}). The potential one-form describing the
electromagnetic field is given by the solution of the source-free
Maxwell equations. In order to construct this solution we shall
use the well-known fact that for a Ricci-flat metric a Killing
1-form field  is closed and co-closed, that is, it can serve as a
potential one-form for an associated test Maxwell field. Since the
Myers-Perry metric is Ricci-flat as well, we can take the
potential one-form field as
\begin{equation}
A= \alpha \, \hat\xi_{(t)}\,\,, \label{potform1}
\end{equation}
where the Killing one-form field $\,\hat
\xi_{(t)}\,=\xi_{(t)\mu}\, d x^{\mu}$ is obtained by lowering the
index of the temporal Killing vector in (\ref{killing}) and
$\,\alpha\,$ is an arbitrary constant parameter. To determine this
parameter we examine the Gauss integral for the electric charge of
the black hole
\begin{equation}
Q= {\frac{1}{A_{N-1}}} \oint \,^{\star}F\,\,, \label{gauss}
\end{equation}
where the $\,{\star}\,$ operator denotes the Hodge dual, along
with expression  (\ref{mass}) for the mass parameter. As a result
we find that
\begin{equation}
\alpha = -\frac{Q}{m\,(N-2)}\,\,. \label{fixpara}
\end{equation}
With this in mind and requiring the vanishing behavior of the
potential at infinity, we obtain the following expression for the
electromagnetic potential one-form
\begin{equation}
A= -\frac{Q}{(N-2)\, r^{N-4}\,\Sigma}\left(dt- a
\sin^2\theta\,d\phi \right)\,\,. \label{potform2}
\end{equation}
Accordingly, the electromagnetic two-form field is given by
\begin{eqnarray}
\label{2form}
 F&=&- \frac{Q}{(N-2)\, r^{N-3}\,\Sigma{\,^2}} \left\{\,
H\,\left(dt-a
\sin^2\theta\,d\phi \right) \wedge dr \right. \\[2mm]  & & \left. \nonumber
 - r a  \sin 2\theta \left[a \,dt - \left(r^2+a^2 \right)\,d\phi
\right] \wedge d\theta  \right\}\,\,,
\end{eqnarray}
where
\begin{equation}
H=(N-2) \,\Sigma -2\, a^2 \cos^2\theta \,\,. \label{h}
\end{equation}
It is also useful to calculate the contravariant components of the
electromagnetic field tensor. They are given by
\begin{eqnarray}
F^{01}&=&\frac{Q}{N-2}\,\,\frac{r^2+a^2}{r^{N-3}\,\Sigma^3}\,H= -
\frac{r^2+a^2}{a}\, F^{13} \,\,,\nonumber\\[2mm]
F^{23}&=& \frac{2 a Q}{N-2}\,\,\frac{\cot\theta}{
r^{N-4}\,\Sigma^3}= - \frac{F^{02}}{a \sin^2{\theta}}\,\,.
\label{contra}
\end{eqnarray}
We recall that equations (\ref{potform2}) and (\ref{2form})
describe the electromagnetic field of a higher dimensional
rotating black hole carrying a weak (test) electric charge. In the
following we shall suppose that the electric charge is no longer
weak that its electromagnetic field essentially affects the
geometry of space-time  around the black hole.

\section{The metric ansatz}

For an arbitrary amount of the electric charge of the black hole
we must solve the simultaneous system of the Einstein-Maxwell
equations. Following the strategy of obtaining the consistent
solution to the Einstein-Maxwell equations in four dimensions
which results in the familiar Kerr-Newman metric, we take the
metric ansatz of the form
\begin{eqnarray}
ds^2 & = &- \left(1-\frac{m}{r^{N-4} \,\Sigma}+
\frac{q^2}{r^{2(N-3)} \,\Sigma} \right) dt^2+ {{r^{\,N-2}\, \Sigma
}\over \Delta}\,dr^2+ \Sigma \, d\theta^2 - \frac{2 a\left(m
r^{N-2}-q^2\right)
\sin^2 \theta}{r^{2(N-3)}\,\Sigma}\,dt \,d\phi \nonumber \\[4mm] &&
 + \left(r^2+a^2+
\frac{a^2 \left(m r^{N-2}-q^2\right) \sin^2\theta}
{r^{2(N-3)}\,\Sigma}\right)\sin^2{\theta}\,d\phi^2  + r^2
\cos^2{\theta} \, d\Omega_{N-3}^2\,\,, \label{ansatz}
\end{eqnarray}
where $\,q\,$ is a parameter related to the physical charge of the
black hole and the metric function $\,\Delta\,$ is now given as
\begin{equation}
\Delta= r^{N-2}(r^2 + a^2) -m \, r^2 + q^2\, r^{4-N}\,\,.
\end{equation}
In choosing this metric ansatz we required the potential one-form
(\ref{potform2}) still to satisfy the Maxwell equations. Indeed,
this idea implicitly  lies at the root of using the metric ansatz
in the Kerr-Schild form , as well as the complex coordinate
transformation method \cite{njanis} to obtain the Kerr-Newman
solution in general relativity. Therefore, the metric ansatz
(\ref{ansatz}) agrees with that suggested in \cite{xu} by applying
the complex coordinate transformation method of \cite{njanis} to a
non-rotating charged  black hole in $\, N+1\,$ dimensions.
However, in our case it is obtained from the Myers-Perry metric
(\ref{metric}) by a simple re-scaling of the mass parameter
\begin{equation}
m\rightarrow m- q^2/r^{N-2}\,\,.\label{resc}
\end{equation}
 Straightforward calculations show that the solution of the source-free Maxwell equations
\begin{equation}
\partial_{\nu} (\sqrt{-g}\,F^{\mu \nu})= 0 \,\, \label{max1}
\end{equation}
is still given by the potential one-form field (\ref{potform2})
and the components of the electromagnetic field tensor given by
(\ref{2form}) and (\ref{contra}) remain unchanged in the metric
(\ref{ansatz}). Therefore, we shall use them to calculate the
energy-momentum source  on the right-hand-side of the higher
dimensional Einstein equations
\begin{equation}
R^{\mu}_{\nu}= 8 \pi G  M^{\mu}_{\nu}\,\, \label{einstein}
\end{equation}
with
\begin{equation}
M^{\mu}_{\nu}=  \frac{1}{A_{N-1}}\left(F^{\mu \alpha} F_{\nu
\alpha}-\frac{1}{2(N-1)}\,\delta^{\mu}_{\nu}\,F_{\alpha \beta}
F^{\alpha \beta}  \right) \,\,. \label{emt}
\end{equation}
The non-vanishing components of the energy-momentum tensor as well
as the Ricci tensor are given in Appendix A and B  by equations
(\ref{sourcecomps1})\,-\,(\ref{sourcecomps6}) and
(\ref{riccicomps1})\,-\,(\ref{riccicomps5}) . It is instructive to
start with the case $\,N=3\,$ . The corresponding energy-momentum
tensor has the components
\begin{eqnarray}
M^{0}_{0}&=& - M^{3}_{3}= -\frac{Q^2}{8 \pi \Sigma^3}
\,\left(r^2+a^2+a^2 \sin^2 \theta \right)\,\,, \nonumber \\[2mm]
M^{1}_{1}&=& -M^{2}_{2}= - \frac{Q^2}{8 \pi \Sigma^2}\,\,,~~~~~
M^{3}_{0}=- \frac{M^{0}_{3}}{\left(r^2+a^2\right) \sin^2 \theta}
=- \frac{a Q^2}{4 \pi \Sigma^3}\,\,, \label{fourd1}
\end{eqnarray}
while, for the components of the Ricci tensor we have
\begin{eqnarray}
R^{0}_{0}&=& - R^{3}_{3}= -\frac{q^2}{\Sigma^3}
\,\left(r^2+a^2+a^2 \sin^2 \theta \right)\,\,, \nonumber \\[2mm]
R^{1}_{1}&=& -R^{2}_{2}= - \frac{q^2}{\Sigma^2}\,\,,~~~~~
R^{3}_{0}=- \frac{R^{0}_{3}}{\left(r^2+a^2\right) \sin^2 \theta}
=- \frac{2\, a q^2}{\Sigma^3}\,\,. \label{fourd2}
\end{eqnarray}
Inspecting equation (\ref{einstein}) with expressions
(\ref{fourd1}) and  (\ref{fourd2}) we see that it is satisfied if
$\,q^2=G Q^2\,.$ This is the case of a rotating charged black hole
in four dimensions and the metric (\ref{ansatz}), with dropped
last term in it, reduces to the Kerr-Newman metric. However, for
the case $\,N \geq 4\,$ equation (\ref{einstein}) is not satisfied
with the expressions (\ref{sourcecomps1})\,-\,(\ref{sourcecomps6})
and (\ref{riccicomps1})\,-\,(\ref{riccicomps5}). It is only
satisfied when we restrict ourselves to a slow rotation of the
black hole. Indeed, to first order in the rotation parameter we
obtain that
\begin{eqnarray}
M^{0}_{0}&=& M^{1}_{1}= -\frac{N-2}{(N-1)\,A_{N-1}}
\,\,\,\frac{Q^2}{r^{2(N-1)}}\,\,,\nonumber \\[2mm]
M^{2}_{2}&=& M^{3}_{3} = M^{4}_{4}= \frac{1}{(N-1)\,A_{N-1}}
\,\,\,\frac{Q^2}{r^{2(N-1)}}\,\,, \\[2mm]
M^{0}_{3}&=-&r^2 \sin^2 \theta \,M^{3}_{0} = \frac{a
\sin^2\theta}{A_{N-1}} \,\,\,\frac{Q^2}{r^{2(N-1)}}\,\,. \nonumber
\label{linemt1}
\end{eqnarray}
We note that all the components $\,M^{i}_{i}\,$  with $\,i \geq
4\,$ are equal to each other. We also have the Ricci components
\begin{eqnarray}
R^{0}_{0}&=& R^{1}_{1}=-\,\,\frac{q^2}{r^{2(N-1)}}\,\,(N-2)^2
\,\,, \nonumber \\[2mm]
R^{2}_{2}&=&R^{3}_{3}=R^{4}_{4}=
\frac{q^2}{r^{2(N-1)}}\,\,(N-2)\,\,,\\[2mm]
R^{0}_{3}&=&  -r^2 \sin^2\theta\,R^{3}_{0}= \,\,\frac{q^2 a
\sin^2\theta}{r^{2(N-1)}}\,\,(N-1)(N-2)\,\, \nonumber
\label{linricci1}
\end{eqnarray}
along with the components  $\,R^{i}_{i}\,$ identical  to each
other for all $\,i \geq 4\,$. Inspecting now equation
(\ref{einstein}) we see that it is satisfied provided that the
charge parameter $\,q\,$ is related to the physical charge of the
black hole as
\begin{eqnarray}
q &=&\pm Q \left[\frac{8 \pi G}{(N-2)\,(N-1)\,
A_{N-1}}\right]^{1/2}  \,\,.\label{physcharge1}
\end{eqnarray}
With this in mind, the  metric of a slowly rotating and charged
black hole in $\,N+1\,$ dimensions can be obtained from
(\ref{ansatz}) by ignoring all terms involving $\,a^2\,$ and
higher powers in $\,a\,$. We arrive at the metric
\begin{eqnarray}
ds^2 & = &- \left(1-\frac{m}{r^{N-2}}+ \frac{q^2}{r^{2(N-2)}}
\right) dt^2+ \left(1-\frac{m}{r^{N-2}}+ \frac{q^2}{r^{2(N-2)}}
\right)^{-1} \,dr^2 \nonumber \\[4mm] &&
- \frac{2 \,a \sin^2 \theta}{r^{N-2}}\,\left(m
-\frac{q^2}{r^{N-2}}\right)\,dt \,d\phi + r^2 \left(d\theta^2
+\sin^2{\theta}\,d\phi^2 + \cos^2{\theta} \,\,
d\Omega_{N-3}^2\right) \,\,. \label{desiredm}
\end{eqnarray}
The potential one-form for associated electromagnetic field must
have the form
\begin{equation}
A= -\frac{Q}{(N-2)\, r^{N-2}}\left(dt- a \sin^2\theta\,d\phi
\right)\,\,. \label{linpotform}
\end{equation}
This metric generalizes the higher dimensional
Schwarzschild-Tangherlini solution for non-rotating charged black
holes in the Einstein-Maxwell gravity to include an arbitrarily
small angular momentum of the black holes. For $\,N=4\,$ this
solution reduces to that given in \cite{aliev}.

\section{Two independent angular momenta}

As we have mentioned above, in the general case, the metric
describing rotating black holes in higher dimensions involves
multiple angular momenta in different orthogonal planes of
rotation. For slow enough rotation one can also include an
electric charge into this metric, keeping only linear in rotation
parameter terms in it and using the re-scaling procedure
determined by equation (\ref{resc}). As an instructive example, we
shall focus on the simplest case with two independent angular
momenta associated with two orthogonal $2$-planes of rotation in
five dimensions. We start with the metric ansatz
\begin{eqnarray}
ds^2 & = & -dt^2 +\Xi\,\left(\frac{r^2}{\Pi}\,dr^2 + d\theta^2
\right) +(r^2 + a^2)\,\sin^2\theta \,d\phi^2
+(r^2 + b^2)\,\cos^2\theta \,d\psi^2 \nonumber \\
&  & +\,\frac{m\,r^2- q^2}{\Xi\,r^2}\, \left(dt - a\, \sin^2\theta
\,d\phi - b\,\cos^2\theta \,d\psi \right)^2\,\,, \label{metric2}
\end{eqnarray}
where the metric functions $\,\Xi\,$ and $\,\Pi\,$ are given as
\begin{equation}
\Xi=r^2+a^2 \,\cos^2\theta + b^2 \,\sin^2 \theta, \;\;\;\;\;\;
\Pi= (r^2 + a^2)(r^2 + b^2) -m \, r^2+q^2 \,.
\end{equation}
One can show that in this metric the electromagnetic field of an
electric charge  $\,Q\,$ is described by the potential one-form
\begin{equation}
A = -\frac{Q}{2\,\Xi}\, \left(d\,t - a\,\sin^2\theta\,d\,\phi -
b\,\cos^2\theta\,d\,\psi\right)\,\,. \label{potform3}
\end{equation}
This is exactly the same potential as that found in \cite{af} for
the unperturbed Myers-Perry metric with a small enough amount of
the electric charge. However, the system of Einstein-Maxwell
equations (\ref{einstein}) and (\ref{max1}) becomes consistent
provided that we  discard all terms with higher powers of rotation
parameters, keeping only linear in $\,a\,$ and $\,b\,$ terms. As a
result, we have the metric
\begin{eqnarray}
ds^2 & = & -\left(1-\frac{m}{r^2}+\frac{q^2}{r^4} \right)\,dt^2
+\left(1-\frac{m}{r^2}+\frac{q^2}{r^4} \right)^{-1}dr^2 +
r^2\left( \,d\theta^2
 + \sin^2\theta\,d\phi^2 + \cos^2\theta \,d\psi^2\right)
\nonumber \\[2mm]
& & -
\frac{2}{r^2}\left(m-\frac{q^2}{r^2}\right)\left(\,a\sin^2\theta
\,d t\,d\phi+ b \cos^2\theta \,dt\, d\psi \right)\,\,,
\label{fmetric}
\end{eqnarray}
where the  parameter $\,q\,$ can be related to the physical charge
of the black hole by inspecting the Einstein equations
(\ref{einstein}) with the associated components of the source
tensor
\begin{eqnarray}
M^{0}_{0}&=& M^{1}_{1}= -\frac{1}{3 \pi^2}
\,\frac{Q^2}{r^{6}}\,\,,~~~~~~~~~~~~~~~~~~~~ M^{2}_{2}= M^{3}_{3}
= M^{4}_{4}\,\,=\, \frac{1}{6 \pi^2}
\,\frac{Q^2}{r^{6}}\,\,, \\[2mm]
M^{0}_{3}&=-&r^2 \sin^2 \theta \,M^{3}_{0}\,\, = \,\,\frac{a
\sin^2\theta}{2 \pi^2} \,\frac{Q^2}{r^{6}}\,\,,~~~~~
M^{0}_{4}=-r^2 \cos^2 \theta \,M^{4}_{0}\,\, = \,\,\frac{b
\cos^2\theta}{2 \pi^2} \,\frac{Q^2}{r^{6}}\,\, \nonumber
\label{linemt2}
\end{eqnarray}
and those of the Ricci tensor
\begin{eqnarray}
R^{0}_{0}&=& R^{1}_{1}=-\,\frac{4
q^2}{r^{6}}\,\,,~~~~~~~~~~~~~~~~~~~~~~~~~~~~~
R^{2}_{2}=R^{3}_{3}=R^{4}_{4}=
\frac{2 q^2}{r^{6}}\,\,,\\[2mm]
R^{0}_{3}&=&  -r^2 \sin^2\theta\,R^{3}_{0}\,\,= \,\,\frac{6\, q^2
a \sin^2\theta}{r^{6}}\,\,,~~~~~~~ R^{0}_{4}= -r^2
\cos^2\theta\,R^{4}_{0}\,\,= \,\,\frac{6\, q^2 b
\cos^2\theta}{r^{6}}\,\, .\nonumber  \label{linricci2}
\end{eqnarray}
We find the relation
\begin{eqnarray}
q &=&\pm Q \sqrt{\frac{2 G}{3 \pi}} \,\,\,\label{physcharge2}
\end{eqnarray}
which, of course, agrees with (\ref{physcharge1}) .

\section{Gyromagnetic Ratio}

It is clear that many of the interesting physical properties of
generic rotating black holes in higher dimensional
Einstein-Maxwell gravity, such as the surface gravity and the
geometry of the event horizon, must crucially depend on $\,a^2\,$
values of the rotation parameter. Fortunately the gyromagnetic
ratio of the black hole can be learned from the limit of slow
rotation described by the metric we discussed above. We recall
that the gyromagnetic ratio is the ratio of the magnetic dipole
moment of a rotating charged black hole to its angular momentum.
One of the remarkable facts about the black hole in the
Einstein-Maxwell theory in four dimensions is that it can be
assigned a gyromagnetic ratio $\,g=2\,$\, just like the electron
in Dirac theory \cite{carter2}, rather than the usual charged
matter in classical electrodynamics, for which $\,g=1\,$. Here we
wish to know how does the value of the gyromagnetic ratio change
for higher dimensional black holes.

From the asymptotic behavior of the metric (\ref{desiredm}) we
find that
\begin{eqnarray}
g_{03}& = & - \,\frac{j\,\sin^2\theta}{r^{N-2}}+
\mathcal{O}\left(\frac{1}{r^{2(N-2)}}\right)\,\,, \label{asympg03}
\end{eqnarray}
which gives the specific angular momentum $\,j \,$ defined in
equation (\ref{momentum}). As for the magnetic dipole moment, it
can also be determined from the far distant behavior of the
magnetic field generated by a rotating and charged black hole
around itself. To describe the magnetic field  in $\,N+1\,$
dimensions it is useful to introduce the magnetic $\,(N-2)\,$-form
defined as
\begin{eqnarray}
{\hat B}_{N-2} &=& i_{\hat\xi_{(t)}}\, ^{\star}F =
^{\star}\left({\hat\xi_{(t)}}\wedge F\right)\,\,, \label{mform}
\end{eqnarray}
which can also be written in the alternative form as follows
\begin{eqnarray}
{\hat B}_{N-2} = \frac{1}{2\,(N-2)\,!}\,\,\sqrt{-g}\,
\,\epsilon_{\,\mu\,\alpha \beta \,i_{1}\,i_{2}\,...\,i_{N-2}}\,
\,\xi_{(t)}^\mu \,F^{\alpha\beta}\, dx^{i_{1}}\wedge
dx^{i_{2}}\wedge...\wedge dx^{i_{N-2}}\,\,. \label{altmform}
\end{eqnarray}
Substituting into this equation the non-vanishing  components of
the electromagnetic field which is given by (\ref{contra}) taken
in the limit of slow rotation, we obtain that
\begin{eqnarray}
{\hat B}_{N-2}&=& \frac{Q
a}{r^2}\,\,\sqrt{\gamma}\,\cos^{\,N-3}{\theta}
\left(\frac{2\cos\theta}{N-2}\,\,\frac{d r}{r} +\sin\theta
\,d\theta \right) \wedge d\chi_{1}\wedge d\chi_{2} \wedge...\wedge
d\chi_{N-3}\,\,. \label{asymmform}
\end{eqnarray}
In the asymptotic rest frame of the black hole the magnetic
$\,(N-2)\,$-form has the following orthonormal components
\begin{eqnarray}
B_{\hat r \,\hat\chi_{1} \hat\chi_2...\hat\chi_{N-3}} &=&
\frac{2\, Q a}{N-2}\,\,\frac{\cos\theta}{r^N}\,\,,~~~~~~~~~B_{\hat
\theta \,\hat\chi_1 \hat\chi_2...\hat\chi_{N-3}} = \frac{Q a
\sin\theta}{r^N}\,\,,
\end{eqnarray}
which is obtained by projecting (\ref{asymmform}) on the basis
(\ref{dual}). These expressions describe the dominant behavior of
the magnetic field far from the black hole and show that the black
hole can be assigned  a magnetic dipole moment given by
\begin{equation}
\mu  = Q\,a \,\,.
\label{dipole}
\end{equation}
We see that the coupling of the rotation parameter of the black
hole to its charge to give the magnetic dipole moment looks
exactly the same as its coupling to the mass parameter to
determine the specific angular momenta $\,j= a m\,$. Thus, we can
write
\begin{eqnarray}
\mu&=&\frac{Q\,j}{m} = (N-1)\,\frac{Q\,J}{2 \,M}\,\, , \label{g1}
\end{eqnarray}
where we have used the relations (\ref{total}) .

Defining now the gyromagnetic parameter  $\,g\,$ in the usual way,
as a constant of proportionality in the equation
\begin{equation}
\mu= g\, \frac{Q\,J}{2 \,M}\,\, \label{g2}
\end{equation}
and comparing this equation with (\ref{g1}) we infer that a
rotating charged black hole in $\, N+1\,$ dimensions possesses the
value of the gyromagnetic ratio
\begin{equation}
g=N-1\,\,. \label{gyro}
\end{equation}
We recall that the surface of the event horizon of the black hole
is topologically equivalent to a $\,(N-1)\,$-sphere. It is
interesting to note that the value of the gyromagnetic ratio
coincides with the dimension of this sphere. In the same way, one
can show that a five-dimensional black hole which is described by
the metric (\ref{fmetric}) can be assigned a gyromagnetic ratio of
value $\,g=3\,$. In \cite{af} we proved that a five dimensional
weakly charged Myers-Perry black holes  must have the value of the
gyromagnetic ratio $\,g=3\,$. We see that in our case this value
remains unchanged for an arbitrary amount of the electric charge.
Moreover, as the specific angular momentum and the magnetic dipole
moment first appear at linear order in rotation, on the grounds of
all described above (see also Ref.\cite{af}), it is
straightforward to verify that the relation
\begin{eqnarray}
\mu_{(i)}&=&\frac{Q\,j_{(i)}}{m} \, \label{multig}
\end{eqnarray}
holds for rotating black holes with multiple independent rotation
parameters $\,a_{(i)}\,$ and accordingly, with multiple magnetic
dipole moments $\,\mu_{(i)}\,$. Since for each orthogonal plane of
rotation the specific angular momentum is given by (\ref{total}),
from equation (\ref{multig}) we conclude that the value of the
gyromagnetic ratio (\ref{gyro}) is also true in the general case
of multiple rotations in $\,N+1\,$ dimensions .

\section{Conclusion}

We have presented a new analytical solution subject to the higher
dimensional Einstein-Maxwell equations which describes rotating
charged black holes with a single angular momentum in the limit of
slow rotation. Earlier, black hole solutions in the Einstein and
Einstein-Maxwell gravity have been discussed by Tangherlini, who
found the exact counterparts of the Schwarzschild and
Reissner-Nordstrom metrics in arbitrary dimensions. In further
developments Myers and Perry have given the exact metric for a
rotating black hole using for it the familiar Kerr-Schild ansatz
in higher dimensions. We have examined the case of a rotating
black hole carrying a Maxwell electric charge. Following the
strategy of obtaining the Kerr-Newman solution in general
relativity we have assumed a special ansatz for the metric of the
rotating charged black hole in arbitrary dimensions. We have shown
that this approach enables one to write down the consistent
solution of the Einstein-Maxwell equations only in the case of
slow rotation. We have also extend this approach to give the
metric for a slowly rotating charged black hole with two
independent angular momenta associated with two orthogonal
$2$-planes of rotation in five dimensions.

Although we could not write down the desired solution for all
values of rotation parameter, the solutions presented in the limit
of slow rotation are sufficient to compute the gyromagnetic ratio
for black holes in higher dimensional Einstein-Maxwell gravity.
Since the angular momenta and the magnetic dipole momenta of these
black holes first appear at the linear order in rotation parameter
we have led to the conclusion that the value of the gyromagnetic
ratio $\,g=N-1\,$ remains the same for multi-rotating and charged
black holes with the spherical topology of the horizon.

\appendix

\section{The Electromagnetic Source Tensor}

Since in higher dimensions the energy-momentum tensor of the
electromagnetic field
\begin{equation}
T^{\mu}_{\nu}=  \frac{1}{A_{N-1}}\left(F^{\mu \alpha} F_{\nu
\alpha}-\frac{1}{4}\,\delta^{\mu}_{\nu}\,F_{\alpha \beta}
F^{\alpha \beta}  \right) \,\, \label{emt1}
\end{equation}
has the non-vanishing trace, the source  on the right-hand-side of
the Einstein equation is given by (\ref{emt}) and it is related to
the canonical form (\ref{emt1}) as
\begin{equation}
M^{\mu}_{\nu}= T^{\mu}_{\nu}- \frac{1}{N-1}\,
\delta^{\mu}_{\nu}\,T \,\,.\label{emt2}
\end{equation}
Using equations (\ref{2form}) and (\ref{contra}) in (\ref{emt}) we
find that the non-vanishing  components of the electromagnetic
source tensor have the form
\begin{eqnarray}
M^{0}_{0}&=&\frac{\lambda^2}{\Sigma^5}\,\,\frac{ \left[\Sigma -
(N-1)(r^2+a^2)\right] H^2 - 4 a^2 r^2 \cos^2\theta
\left[\Sigma+(N-1) \,a^2 \sin^2\theta\right]}{\,r^{2(N-3)}}\,\,,
\label{sourcecomps1}
\end{eqnarray}
\begin{eqnarray}
M^{3}_{3}&=&\frac{\lambda^2}{\Sigma^5}\,\,\frac{ \left[\Sigma +
(N-1)\,a^2 \sin^2\theta \right] H^2 - 4\, a^2 r^2 \cos^2\theta
\left[\Sigma-(N-1) \,(r^2+a^2)\right]}{r^{2(N-3)}} \,\,,
\label{sourcecomps2}
\end{eqnarray}
\begin{eqnarray}
M^{3}_{0}&=-& \frac{M^{0}_{3}}{(r^2+a^2) \sin^2\theta}=-
\frac{\lambda^2\,a}{\Sigma^4}\,\,\frac{(N-1) \left[(N-2)^2\,\Sigma
- 4\,a^2 (N-3) \cos^2\theta \right]}{r^{2(N-3)}}
\,\,,\label{sourcecomps3}
\end{eqnarray}
\begin{eqnarray}
M^{1}_{1}&=-&\frac{\lambda^2}{\Sigma^4}\,\,\frac{ (N-2)\, H^2 +4\,
a^2 r^2 \cos^2 \theta}{r^{2(N-3)}} \,\,,\label{sourcecomps4}
\end{eqnarray}
\begin{eqnarray}
M^{2}_{2}&=&\frac{\lambda^2}{\Sigma^4}\,\,\frac{ H^2 +4\, a^2 r^2
(N-2) \cos^2 \theta}{r^{2(N-3)}} \,\,, \label{sourcecomps5}
\end{eqnarray}
\begin{eqnarray}
M^{4}_{4}&=&\frac{\lambda^2}{\Sigma^4}\,\,\frac{ H^2 -4\, a^2 r^2
\cos^2 \theta}{r^{2(N-3)}} \,\,, \label{sourcecomps6}
\end{eqnarray}
and all $\, M^{i}_{i}\,$ with $\,i \geq 4\,$ become equal to each
other. In these expressions the function $\,H\,$ is given by
(\ref{h}) and we have also introduced the notation
\begin{eqnarray}
\lambda^2 &=&\frac{Q^2}{(N-2)^2\,(N-1)\,
A_{N-1}}\,\,.\label{notation}
\end{eqnarray}
We note that the relation
\begin{eqnarray}
M^{0}_{0} + M^{3}_{3}&=& M^{1}_{1} + M^{2}_{2}=-
(N-3)\,M^{4}_{4}\,\,\label{id1}
\end{eqnarray}
holds between the components of the energy-momentum tensor .
\begin{center}
\section{ Calculating the components of the Ricci tensor}
\end{center}

In order to calculate the non-vanishing components of the Ricci
tensor it is convenient to use the method of orthonormal basis
forms. The basis one-forms for the metric (\ref{ansatz}) can be
chosen as
\begin{eqnarray}
\omega\,^{\hat{0}} & =& \left| g_{00}- \omega^2\,g_{33}\right|^{1/2}\,d\,t \nonumber\\[2mm]
\omega\,^{\hat{1}} & =&  (\,g_{11}\,)\,^{1/2}\,d\,r \nonumber\\[2mm]
\omega\,^{\hat{2}} & =& (\,g_{22}\,)\,^{1/2}\,d\,\theta \,\,,\nonumber\\[2mm]
\omega\,^{\hat{3}} & =& \left(g_{33}\right)^{\,1/2}
\,\left(d\,\phi- \omega \,d\,t\right)
\nonumber\\[2mm]
\omega\,^{\hat{4}} & =& r \cos\theta\,d\,\chi_1 \,\,,\nonumber\\[2mm]
\omega\,^{\hat{5}} & =& r \cos\theta\,\sin\chi_1 d\,\chi_2 \,\,,\nonumber\\
\vdots&  =&\vdots
 \label{frame}
\end{eqnarray}
where we have introduced the "angular velocity" of rotation
\begin{equation}
\omega=-\,\frac{g_{03}}{g_{33}}
\end{equation}
The corresponding dual basis is given by
\begin{eqnarray}
e\,_{\hat{0}} & =& \left| g_{00}-
\omega^2\,g_{33}\right|^{-1/2}\,\left(\frac{\partial}{\partial t}
 + \omega\, \frac{\partial}{\partial \phi}\right)\nonumber\\[2mm]
e\,_{\hat{1}} & =&
\frac{1}{(\,g_{11}\,)\,^{1/2}}\,\frac{\partial}{\partial r}
\nonumber\\[2mm]
e\,_{\hat{2}} & =&
\frac{1}{(\,g_{22}\,)\,^{1/2}}\,\frac{\partial}{\partial \theta}
 \,\,,\nonumber\\[2mm]
e\,_{\hat{3}} & =&\frac{1}{(\,g_{33}\,)\,^{1/2}}\,\frac{\partial}{\partial \phi}\,\,,\nonumber\\[2mm]
e\,_{\hat{4}} & =& \frac{1}{r \cos\theta}\,\frac{\partial}{\partial \chi_1}\,\,,\nonumber\\[2mm]
e\,_{\hat{5}} & =& \frac{1}{r \cos\theta
\sin\chi_1}\,\frac{\partial}{\partial \chi_2}\,\,,\nonumber\\
\vdots&  =&\vdots
\label{dual}
\end{eqnarray}
We calculate the anti-symmetric connection one-forms and the
components of the Riemann tensor through the Cartan's  structure
equations
\begin{eqnarray}
d\omega^{\hat\mu}+{\omega^{\hat\mu}}_{\hat\nu}\wedge
\omega^{\hat\nu}&=&0\,\,,~~~~~~~ {{\cal
R}^{\hat\mu}}_{\hat\nu}=d{\omega^{\hat\mu}}_{\hat\nu}+{\omega^{\hat\mu}}_{\hat\alpha}\wedge
{\omega^{\hat\alpha}}_{\hat\nu}\,\,,
\end{eqnarray}
where the curvature 2-form is defined as
\begin{equation}
{\cal R}^{\hat\mu\hat\nu}=\frac{1}{2}
{R^{\hat\mu\hat\nu}}_{\hat\alpha
\hat\beta}\,\omega^{\hat\alpha}\wedge \omega^{\hat\beta}\,\,.
\end{equation}
Next, after straightforward calculations we find that the
non-vanishing components of the Ricci tensor in the coordinate
basis  are given by
\begin{eqnarray}
R^{0}_{0}&=-&\frac{q^2\,(N-2)}{\Sigma^3}\,\,\frac{
\left[(N-2)(r^2+a^2)-a^2 \right]\Sigma + 2\, a^2 r^2
\sin^2\theta}{r^{2(N-2)}}\,\,,\label{riccicomps1}
\end{eqnarray}
\begin{eqnarray}
R^{3}_{3}&=&\frac{q^2\,(N-2)} {4\,\Sigma^3}\, \,\,\frac{(N-3)\,
a^4 \sin^2 2\,\theta +4\,r^2\left[\Sigma + (N-1)\,a^2 \sin^2\theta
\right] }{r^{2(N-2)}}\,\,,\label{riccicomps2}
\end{eqnarray}
\begin{eqnarray}
R^{3}_{0}&=-& \frac{R^{0}_{3}}{(r^2+a^2) \sin^2\theta}=-
\,\,\frac{q^2 a (N-2)}{\Sigma^3}\,\,\frac{(N-1)\,\Sigma - 2\,a^2
\cos^2\theta}{r^{2(N-2)}} \,\,,\label{riccicomps3}
\end{eqnarray}
\begin{eqnarray}
R^{1}_{1}&=&- \,\,\frac{q^2
(N-2)}{\Sigma^2}\,\,\frac{(N-2)\,\Sigma - \,a^2
\cos^2\theta}{r^{2(N-2)}}
\,\,,\;\;\;\;\;\;R^{2}_{2}=R^{4}_{4}\,\frac{r^2}{\Sigma}\,\,,\label{riccicomps4}
\end{eqnarray}
\begin{eqnarray}
R^{4}_{4}&=&
\,\,\frac{q^2}{\Sigma}\,\,\,\frac{N-2}{r^{2(N-2)}}\,\,
\label{riccicomps5}
\end{eqnarray}
together with the components $\, R^{i}_{i}\,$ which are the same
for all $\,i \geq 4\,$. Furthermore, one can show that similar to
(\ref{id1}), the components of the Ricci tensor obey the relation
\begin{eqnarray}
R^{0}_{0} + R^{3}_{3}&=& R^{1}_{1} + R^{2}_{2}=-(N-3)\,R^{4}_{4}
\,\,.\label{id2}
\end{eqnarray}

\end{document}